\documentclass{appolb}
\usepackage{amsmath}
\usepackage{graphicx}

\begin{document}
\title{Model independent analysis of nearly L\'evy correlations%
\thanks{Presented at 11th Workshop on Particle Correlations and
      Femtoscopy (WPCF 2015), WUT, Warsaw, Poland, November 3-7, 2015}%
}
\author{T. Nov\'ak$^{1}$, T. Cs\"org\H{o}$^{1,2}$, H.C. Eggers$^3$ and M. de Kock$^3$
\address{
$^1$KRF, H-3200 Gy\"ongy\"os, M\'atrai \'ut 36, Hungary, \\
$^2$Wigner RCP, H--1525 Budapest 114, POBox 49, Hungary, \\
$^3$Dept.\ of Physics, Stellenbosch University, ZA--7600 Stellenbosch, South Africa
}
}

\maketitle
\begin{abstract}
  A model-independent method for the analysis of two-particle
  short-range correlations is presented. It can be utilized to
  describe such Bose-Einstein (HBT), dynamical (ridge) and other
  correlation functions which have a nearly L\'evy or ``stretched
  exponential'' shape.  For the special case of L\'evy exponent
  $\alpha = 1$, earlier Laguerre expansions are recovered, while for
  $\alpha = 2$ a new expansion is obtained for correlations which are
  nearly Gaussian in shape.  Multidimensional L\'evy expansions are
  also introduced and their potential application to analyze ridge
  correlation data is discussed.
\end{abstract}

  
\section{Introduction}

The detailed shape analysis of the two-particle Bose-Einsten
Correlations (BEC) is an important topic in high energy particle and
nuclear physics because the shape of the correlation function carries
information about the space-time structure of the particle emission
process~\cite{Kittel:2001zw, Csorgo:1999sj}.

With a few assumptions~\cite{Csorgo:1999sj} the two-particle
correlation function is related to the Fourier transformed source
distribution. In this article, however, we do not assume such a
relationship between Fourier transformed source distributions, and
measured two-particle correlations, because in some cases these
assumptions are experimentally shown to be invalid even if they were
expected to hold~\cite{Achard:2011zza}. Instead, we continue the
development of a {\it model-independent} method of analyzing
short-range correlations, developing further the ideas suggested in
Ref.~\cite{Csorgo:2000pf}.  This general, model-independent
characterization of short-range correlation
functions~\cite{Csorgo:2000pf} depends on the following experimental
conditions:
\begin{description}
\item {\it (i)} The correlation function tends to a constant for large
  values of the relative momentum $Q$.
\item {\it (ii)} The correlation function has a non-trivial structure
  at a certain value of its argument, here assumed to be around
  $Q = 0$.
\end{description}
The first applications of this method investigated nearly Gaussian and
nearly exponential correlations~\cite{Csorgo:2000pf,Csorgo:2003uv}.
Here we continue the investigations started in
Ref.~\cite{DeKock:2012gp}, to develop a model-independent technique to
analyze short-range correlations that have a stretched exponential or
L\'evy shape in zeroth order approximation; hence we add a third
experimental precondition:
\begin{description}
\item {\it (iii)} The short-range behaviour of the correlation
  function has a form which is close to the stretched exponential
  i.e.\ an exponential in the stretched variable $Q^{\alpha}$ with
  $0\leq \alpha\leq 2$.
\end{description}
We compare the resulting L\'evy expansion series to the earlier
results for the $\alpha =$ 1 and 2 special cases, and extend this
analysis in a natural manner to the case of multivariate, nearly
symmetric L\'evy distributions.

\section{Univariate L\'evy expansions}

In order to characterize the deviation of the correlation shape from
the approximate L\'evy shape, we apply the general expansion method of
Ref.~\cite{Csorgo:2000pf} for the special case of the L\'evy weight
function, $t = Q R$,
$w(t|\alpha) = \exp(-t^{\alpha}) = \exp(-Q^{\alpha} R^{\alpha})$ . The
expansion is based on a set of polynomials which are orthonormal with
respect to the weight function $w(t|\alpha)$.  This expansion is
uniquely defined by a Gram-Schmidt process if the order $n$ terms are
order $n$ polynomials, with convergence criteria specified in
Ref.~\cite{Csorgo:2000pf}.

The L\'evy expansion of short range correlation functions results in
the following formula which can be easily fitted to a given data set
as
\begin{eqnarray}
  t  & = & Q R, \\
  C_2(t) & = & N \left\{ 1+\lambda \exp(-t^\alpha) \sum_{n=0}^{\infty} c_n L_n (t|\alpha) \right\},
  \label{e:Levy-expansion}
\end{eqnarray}
where $N$ is a normalization coefficient, $\lambda$ measures the
strength of the correlation function, $\exp(-t^{\alpha})$ is the
weight function and zeroth order approximation for the experimentally
measured correlation function and $\alpha$ is the L\'evy index of
stability.  The expansion coefficients are denoted by $c_n$ and
$\{L_n(t|\alpha) \}_{n=0}^{\infty}$ denote the L\'evy polynomials, a
complete set of polynomials which are orthogonal with respect to the
L\'evy weight function $\exp(-t^{\alpha})$.  These L\'evy polynomials
were introduced in Ref.~~\cite{DeKock:2012gp}; the first three are
\begin{eqnarray}
  L_0(t\,|\,\alpha) & =&  1 , \\
  L_1(t\,|\,\alpha) & =& 
	\det\left(\begin{array}{c@{\hspace*{8pt}}c}
      \mu_{0,\alpha} & \mu_{1,\alpha} \nonumber \\ 
      1 & t \end{array} \right) , \\
  L_2(t\,|\,\alpha) & =& 
	\det\left(\begin{array}{c@{\hspace*{8pt}}c@{\hspace*{8pt}}c}
      \mu_{0,\alpha} & \mu_{1,\alpha} & \mu_{2,\alpha} \\ 
      \mu_{1,\alpha} & \mu_{2,\alpha} & \mu_{3,\alpha} \nonumber \\ 
      1 & t & t^2 \end{array} \right) \qquad \text{etc.}
\end{eqnarray}
where 
$$
\mu_{n,\alpha} = \int_0^\infty dt\;t^{n} \exp( - t^\alpha) =
\tfrac{1}{\alpha}\,\Gamma( \tfrac{n+1}{\alpha}) \\
$$
and Euler's gamma function is defined as
$
\Gamma(z) = \int_0^\infty dt\;t^{z-1}e^{-t}
$ as usual.
The lowest-order Levy polynomials are
\begin{eqnarray} 
  L_0(t\,|\,\alpha)  & = & 1
			, \\
  L_1(t\,|\,\alpha)  & = & 
		\frac{1}{\alpha}\left\{
				\Gamma\left(\frac{1}{\alpha}\right) t -
				 \Gamma\left(\frac{2}{\alpha}\right)
		\right\} 
			, \\
  L_2(t\,|\,\alpha) & = & 
		\frac{1}{\alpha^2}\left\{
			\left[
				\Gamma\left(\frac{1}{\alpha}\right) \Gamma\left(\frac{3}{\alpha}\right)
				- \Gamma^2\left(\frac{2}{\alpha}\right)
			\right] t^2 - \right. \nonumber \\ 
			&  & - \left[
				\Gamma\left(\frac{1}{\alpha}\right) \Gamma\left(\frac{4}{\alpha}\right) 
				- \Gamma\left(\frac{3}{\alpha}\right)
				 \Gamma\left(\frac{2}{\alpha}\right)
			\right] t + \\
			&  & \left. + \left[
				\Gamma\left(\frac{2}{\alpha}\right) \Gamma\left(\frac{4}{\alpha}\right) 
				- \Gamma^2\left(\frac{3}{\alpha}\right)
			\right] 
		\right\} . \nonumber
\end{eqnarray}
For $\alpha=1$, these reduce to Laguerre polynomials and the L\'evy
expansion reduces to the Laguerre expansion of
Ref.~\cite{Csorgo:2000pf},
\begin{eqnarray}
  L_0(t\,|\,\alpha = 1) & = & 1 , \\
  L_1(t\,|\,\alpha = 1) & = & t - 1 , \\
  L_2(t\,|\,\alpha = 1) & = & t^2 - 4 t + 2 .
\end{eqnarray}
The $\alpha = 2 $ case provides a new expansion around a Gaussian
shape that is defined for non-negative values of $t$ only,
\begin{eqnarray}
  L_0(t\,|\,\alpha = 2) & = & 
		1, \\
  L_1(t\,|\,\alpha = 2) & = & 
		\frac{1}{2}\left\{
			\sqrt{\pi} t - 1 
		\right\} , \\
  L_2(t\,|\,\alpha = 2) & = & 
		\frac{1}{16}\left\{
			2(\pi - 2) t^2 - 2 \sqrt{\pi} t + (4 - \pi)
		\right\} . 
\end{eqnarray}

\section{Multi-variate Levy expansions}

Multi-variate short range correlations, in particular Bose-Einstein or
HBT correlations, as well as dynamical ``ridge" correlations are
frequently studied in high energy particle and heavy ion physics.  If
they are symmetric ~\cite{Nolan:2016st,Csorgo:2003uv}, a new variable
can be introduced that reduces the multi-variate problem to an
effective one-dimensional problem.
For the multi-variate Bose-Einstein or HBT correlation measurements, a
dimensionless scaling variable has already been
introduced~\cite{Csorgo:2003uv} as follows,
\begin{equation}
t=\left(\sum_{i,j= {\mbox{\rm\tiny side,out,long}}} R_{i,j}^2 q_i q_j \right)^{1/2},
\end{equation}
where $q_i$ stands for the relative momentum component in the given
direction.
For multi-variate angular or ridge correlation measurements that result in a nearly L\'evy 
shape~\cite{Janik:2014hkt}, a similar, dimensionless scaling variable can
be introduced:
\begin{equation}
t=\left(\sum_{i,j={\eta,\phi}} \sigma_{i,j}^2 {\Delta}_i {\Delta}_j \right)^{1/2},
\end{equation}
where $\Delta_i$ stands for the angular difference in the
$(\eta,\phi)$ lego plot.
In both cases, the fit function for a multi-variate, nearly Levy
correlation functions becomes identical to
Eq.~(\ref{e:Levy-expansion}).

\section*{Acknowledgements}
\noindent
We are grateful to the organizers of WPCF 2015 for an inspiring  and useful meeting.
This work was supported in part by the Hungarian OTKA grant NK--101438 and the 
South African National Research Foundation.


\begin{thebibliography}{99}

\bibitem{Kittel:2001zw}
  W.~Kittel,
  Acta Phys.\ Polon.\ B {\bf 32} (2001) 3927.

 
\bibitem{Csorgo:1999sj}
  T.~Cs\"org\H{o},
  Heavy Ion Phys.\  {\bf 15} (2002) 1.

   
\bibitem{Achard:2011zza}
  P.~Achard {\it et al.} [L3 Collaboration],
  Eur.\ Phys.\ J.\ C {\bf 71} (2011) 1648.

   
\bibitem{Csorgo:2000pf}
  T.~Cs\"org\H{o} and S.~Hegyi, %
  Phys.\ Lett.\ B \textbf{489} (2000) 15.

\bibitem{Csorgo:2003uv}
  T.~Cs\"org\H{o}, S.~Hegyi and W.~A.~Zajc,
  Eur.\ Phys.\ J.\ C {\bf 36} (2004) 67.

\bibitem{DeKock:2012gp}
  M.~B.~De Kock, H.~C.~Eggers and T.~Cs\"org\H{o},
  PoS WPCF {\bf 2011} (2011) 033.

\bibitem{Nolan:2016st}
  J.~P.~Nolan,  {\it Stable distributions: Models for Heavy Tailed Data},
  Springer-Verlag, Imprint Birkhauser,  ISBN10 0817641599, 
	(2016), pp.~1-352.
  
\bibitem{Janik:2014hkt}
  M.~Janik, PhD Thesis,
  CERN-THESIS-2014-339.

\end{thebibliography}
\end{document}